# Laterally defined freely suspended quantum dots in GaAs/AlGaAs heterostructures

C Rossler[1], M Bichler[2], D Schuh[3], W Wegscheider[3] and S Ludwig[1]

[1] Center for NanoScience and Fakultät für Physik der Ludwig-Maximilians-Universität, Geschwister-Scholl-Platz 1, 80539 Munich, Germany

[2] Walter Schottky Institut, Technical University Munich, Am Coulombwall 3, 85748 Garching, Germany

[3] Institut für Angewandte und Experimentelle Physik II, University Regensburg, Universitätsstrasse 31, 93040 Regensburg, Germany

email: roessler@lmu.de

**Abstract**

Free standing beams containing a two-dimensional electron system are shaped from a GaAs/AlGaAs heterostructure. Quantum point contacts and (double) quantum dots are laterally defined using metal top gates. We investigate the electronic properties of these nanostructures by transport spectroscopy. Tunable localized electron states in freely suspended nanostructures are a promising tool to investigate the electron-phonon-interaction.

## Introduction

The coherent dynamics of charge states in solid-state-based quantum dots (QDs) is limited at typical experimental temperatures by the electron-phonon interaction [1]. The phonon emission spectrum of a detuned double QD in bulk GaAs has been investigated by measuring the inelastic tunnel current [2]. Double QDs that are defined in a freely suspended bridge are subject to a phonon spectrum that is modified by the cavity modes of the bridge. Therefore, the inelastic tunnel current of a freely suspended double QD will allow a systematic study of the cavity phonon spectrum and its interaction with the electronic states of the QDs. The fabrication of a freely suspended bridge containing metal top gates designed to define a double QD has been reported [3]. A coherent interaction between cavity phonon modes and localized electron states in a QD defined by etched narrow constrictions in a freely suspended beam has been investigated [4]. Here we report on systematic studies of the electronic transport properties of laterally defined suspended QDs. In these devices the charge states are widely tuneable by using metal gates. In particular, we study the influence of metal top gates onto the electronic stability of freely suspended devices.

## Experimental details

The GaAs/Al$_{0.3}$Ga$_{0.7}$As heterostructures (active layer) of thickness 130 nm (sample I) or 90 nm (sample II and sample III) are separated from the substrate by an AlAs/GaAs superlattice (sacrificial layer). In both heterostructures, the active layer contains a two-dimensional electron gas (2DEG) about 40 nm below the surface [5]. The ohmic contacts situated at non-suspended areas allow electric connection of the 2DEG. Optical lithography is used to deposit a 100 nm thick layer of nickel to protect the areas of the heterostructure that will later form the 2DEG-leads. A 60 nm thick layer of gold is deposited for the contacts of the top gates and alignment marks. In the next step 30 nm thick top gates (5 nm titanium, 25 nm gold) are defined by e-beam lithography. In a second step of e-beam lithography, the intended beam area and its connection to the 2DEG-leads is covered by a 100 nm thick layer of nickel. This is done in the last step of lithography because the beam has to be aligned accurately with respect to the gates and the nickel will be removed afterwards. Anisotropic reactive ion etching using SiCl$_4$ removes the uncovered top layers of the heterostructure [6]. The nickel layer is then removed by Fe-III-chloride. Finally, isotropic wet-etching with 1% hydrofluoric acid selectively removes the 400 nm thick sacrificial layer in the vicinity of the already etched trenches. The remaining free-standing beams are about 6 μm long and between 500 nm and 800 nm wide.

One of our devices (sample I) is shown in the atomic force micrograph in Figure 1a. The split gate geometry of these gates was designed to laterally define a double QD. The ten vertical stripes (five on each side of the beam) are metal top gates. They are freely suspended between the beam (dark horizontal stripe) and the sockets (dark rectangles at the top and bottom). The scanning electron micrograph of sample II is shown in Figure 1b. This image has been recorded under a tilt angle of 85° to better visualize the freely suspended regions.

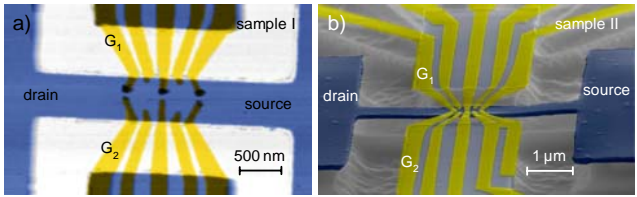

Figure 1. (color online) a) Atomic force micrograph of the split gate device sample I. Ten gates (vertical stripes) lie on top of the freely suspended beam (dark horizontal bar), which is 130 nm thick, 600 nm wide and 6 μm long. b) Scanning electron micrograph of a freely suspended beam (sample II) recorded at a tilt angle of 85°.

The samples were cooled down in a dilution refrigerator with a base temperature of $T_{MC} \cong 20\,\text{mK}$. The electron temperature has been determined as $T_{el} \leq 60$ mK from Coulomb blockade oscillations of non-suspended QDs on a similar sample. After initial illumination with an infrared LED, two–terminal differential conductance measurements through the beams were performed at a frequency of $f = 67$ Hz and modulation voltage of $V = 15\,\mu\text{V}$. The quantized conductance of a quantum point contact (QPC) as a function of identical negative voltages applied to gates $G_1$ and $G_2$ (Figure 1a) is presented in Figure 2a. In these measurements all other gates are grounded. The lead resistance $R_l = 2850\ \Omega$ is already subtracted ($1/G = 1/G_{measured} - R_l$). The value of $R_l$ (measured two-terminal resistance $R_{2T} \cong 3\,\text{k}\Omega$) has been chosen so that the conductance plateaus of the first depletion curve (upper curve in Figure 2a) result to be quantized in multiples of 2e²/h.

**Results and discussion**

While depleting the QPC for the first time (upper curve in Figure 2a) it shows clear conductance plateaus at $G = 2n\,\text{e}^2/h$; n = 0,1,2,…,11. A second depletion process (after slowly sweeping the gate voltages back to zero) results in less pronounced conductance plateaus showing irregular shapes and full depletion at a higher gate voltage. Note, that the lead resistance is almost unaltered compared to the first depletion curve as the plateaus are still close to $G = 2n\,\text{e}^2/h$. The lowest curve in Figure 2a was measured five days later after many depletion cycles. Now the QPC is completely closed at even higher gate voltages. The depletion curve shows characteristic resonances (local extrema) pointing to localized electron states [7].

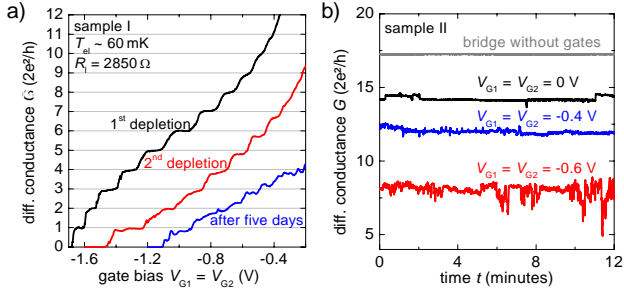

Figure 2. a) Two terminal differential conductance after subtraction of the lead resistance as a function of the voltage applied to gates $G_1$ and $G_2$ while all other gates are grounded. The three traces are measured at subsequent times after illumination. b) Differential conductance as a function of time for an ungated reference bridge (top curve) and for different voltages applied to gates $G_1$ and $G_2$.

Such a behaviour can be explained by electrons that tunnel from the gates into the active layer. A tunnel current between the 2DEG and top gates has previously been reported in a non-suspended heterojunction [8]. Charges tunneling from the surface into the active layer produce reorganizations of the local electrostatic potential causing temporal fluctuations of the conductance as well as permanent depletion. The time dependence of the conductance is shown in Figure 2b for different gate voltages. The top curve is measured on a suspended beam without gates. It shows no pronounced conductance fluctuations during the time period of 12 minutes. The lower three curves are measured on the device shown in Figure 1b. Fluctuations are more frequent at lower gate voltages. This observation supports the model of electrons tunneling from the gates into the active layer. For a more negative gate voltage the electric field between the gates and the 2DEG increases and thereby decreases the tunnel barrier allowing a larger tunneling current of electrons into the active layer [8]. Note, that the amplitude of the conductance fluctuations shown in Figure 2b depends on the strongly varying slope of the depletion curve and is a non monotonous function of the gate voltage. We were not able to define a double QD with this device because the gate voltages necessary to create tunnel barriers result in critical leakage currents and an unstable background potential.

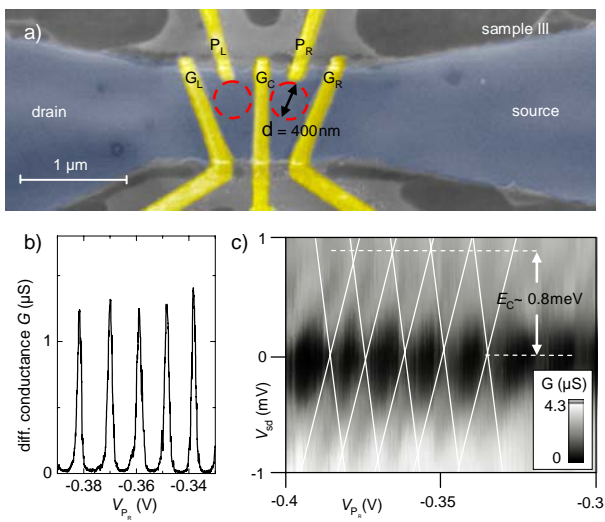

Figure 3. (color online) a) Scanning electron micrograph of sample III (top view). The free-standing beam is 4 μm long, 800 nm wide and 90 nm thick. Three gates $G_L$, $G_C$ and $G_R$ define quantum dots (dashed circles) and two gates $P_L$ and $P_R$ are used to control the number of electrons in the quantum dot. b) Coulomb oscillations of the right QD as a function of $V_{PR}$ for $V_{GC} = V_{GR}$ = -0.275 V c) Gray-scale plot of the differential conductance (linear scale) as a function of $V_{PR}$ and source-drain-bias $V_{SD}$. White lines are a guide for the eyes.

A revised gate design is depicted in Figure 3a. In contrast to the other two samples three gates $G_L$, $G_C$ and $G_R$ completely cross the bridge. The gates define tunnel barriers of the two serial QDs (dashed circles). Two gates $P_L$ and $P_R$ (plunger gates) control the number of electrons in the QDs. The advantage of gates crossing the whole beam (Figure 3a) compared to the split gate design used before (Figure 1) is, that less negative voltages are needed for depletion. Therefore, we expect smaller leakage currents for the Schottky gates and a more stable device. We find that a QD is formed at a voltage $V_G = -0.275\,\text{V}$ applied to gates $G_C$ and $G_R$, where the conductance exhibits Coulomb blockade oscillations as a function of the voltage applied to gate $P_R$ (see Figure 3b). In Figure 3c the conductance is plotted using a linear gray scale as a function of the plunger gate voltage $V_{PR}$ and the source-drain-voltage $V_{SD}$. Coulomb blockade appears as darker regions (lower conductance) in the transport spectrum. From the charging energy $E_c \sim 0.8\,\text{meV}$ and the dielectric constant of GaAs $\varepsilon_{GaAs} \sim 13$ we estimate the diameter of the QD to be $e^2/4\varepsilon_{GaAs}\varepsilon_0 E_c \sim d \sim 400\,\text{nm}$. A very similar QD could be defined between gates $G_L$ and $G_C$ while grounding $G_R$ (data not shown). The estimated diameters of the QDs are in excellent agreement with the sample geometry (Figure 3a). The gate voltage at which each Coulomb blockade oscillation occurs, varies over time by $\Delta V_G \sim \pm 1\,\text{meV}$. This corresponds to a change of the local electrostatic potential of $\Delta\phi \sim 400\,\mu\text{eV}$ which is larger than the phonon cavity energies of $\varepsilon_{ph} \sim 100\,\mu\text{eV}$ expected from numerical calculations [9]. Thus, quantized mechanical excitations can not yet be resolved.

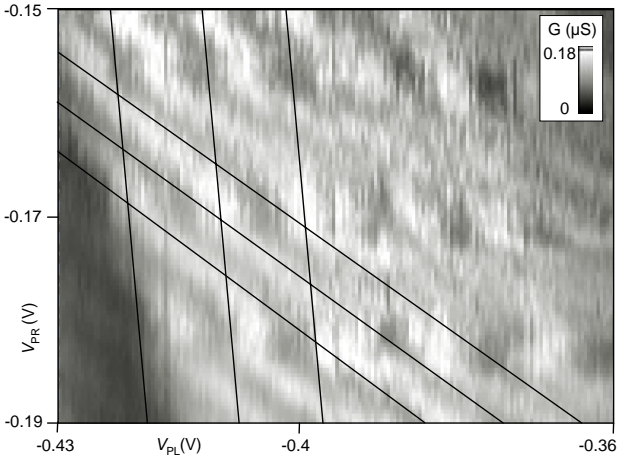

Figure 4. Linear gray-scale plot of the differential conductance of the double QD (Fig. 3a) as a function of the voltages applied to gates $P_L$ and $P_R$. The voltages applied to the other gates are $V_{GL}$ = -0.24 V, $V_{GC}$ = -0.23 V and $V_{GR}$ = -0.23 V.

Figure 4 shows the conductance as a function of both $V_{PL}$ and $V_{PR}$ while applying voltages $V_{GL} = -0.24\,\text{V}$, $V_{GC} = -0.23\,\text{V}$ and $V_{GR} = -0.23\,\text{V}$ to the tunnel barrier gates. Coulomb blockade appears as dark regions (low conductance) in the energy diagram. Bright lines of high conductance exhibit a nonlinear dependence on both plunger gates, typical for the stability diagram of a double QD [10]. Black lines emphasize the two apparent slopes of these resonances, each corresponding to the charging line of one of the two QDs. The fact that the triple points (where the bright lines of two slopes meet) are connected by (bright) lines of high differential conductance indicates that the two QDs are strongly tunnel coupled [11]. The characteristics of a weakly coupled DQD were not observable, probably because the resulting lower tunnel current has been obscured by the apparent conductance fluctuations.

**Conclusion**

In summary, we have investigated freely suspended quantum point contacts and (double) quantum dots defined by split gates. Our studies reveal that the fabricated top gated devices lack the electronic stability needed to employ them for detailed investigations of electronic transport through single and coupled quantum dots on suspended beams. Fluctuations induced at larger gate bias are most likely caused by tunneling of electrons from the gate into the 2DEG. A revised design employed gates that completely cross the bridge in order to reduce the tunnel currents and was successfully used to define a double quantum dot. Still, the fluctuations apparent in this system prevent a systematic investigation of electron-phonon interaction. In future, etched side gates might provide a way to avoid switching noise due to leakage currents while maintaining full tunability of the system.

**Acknowledgments**

The authors thank Stephan Manus and Alexander Paul for technical support and Jorg P. Kotthaus for fruitful discussions. We gratefully acknowledge financial support by the Bundesministerium für Bildung und Forschung via nanoQUIT and the German excellence initiative via the cluster "Nanosystems Initiative Munich (NIM)".